\documentclass[10pt,twocolumn,showpacs,superscriptaddress,notitlepage,prl,aps,amsmath,amssymb]{revtex4}
\usepackage{epsfig}

\begin{document}
\title{Geometrical Frustration and Static Correlations in a Simple Glass Former}
\author{Benoit~Charbonneau}
\affiliation{Mathematics Department, St.Jerome's University in the University of Waterloo, Waterloo, Ontario, Canada}
\author{Patrick~Charbonneau}
\affiliation{Departments of Chemistry and Physics, Duke University, Durham,
North Carolina 27708, USA}
\affiliation{LPTMC, CNRS-UMR 7600, Universit\'e Pierre et Marie Curie, bo\^ite 121,
4 Place Jussieu, 75005 Paris, 
France}
\author{Gilles~Tarjus}
\affiliation{LPTMC, CNRS-UMR 7600, Universit\'e Pierre et Marie Curie, bo\^ite 121,
4 Place Jussieu, 75005 Paris, 
France}

\date{November 7, 2011}

\begin{abstract}
We study the geometrical frustration scenario of glass formation for simple hard sphere models. We find that the dual picture in terms of defects brings little insight and no theoretical simplification for the understanding of the slowing down of relaxation, because of the strong frustration characterizing these systems. The possibility of a growing static length is furthermore found to be physically irrelevant in the regime that is accessible to computer simulations.
\end{abstract}


\maketitle

The ubiquitous glass formation and jamming still puzzle physicists. How can molecular and colloidal systems slow down so abruptly without obvious structural changes? In response to this conundrum, theoretical approaches inspired by spin-glass physics have long postulated a role for a ``hidden'' static length associated with the dynamical slowdown. Following the rigorous identification of a growing point-to-set length accompanying diverging relaxation times in structural glasses~\cite{bouchaud:2004,montanari:2006,franz:2011}, a series of ``order agnostic'' proposals for static correlations in supercooled liquids have also recently flourished~\cite{biroli:2008,karmakar:2009,sausset:2011}, and their analysis is ongoing. Yet specifying a relevant amorphous order parameter that captures these materials' rich phenomenology while providing geometric insights into the underlying microscopic mechanism is still sought after. A scenario for growing geometrical order proposed some time ago by Sadoc and Mosseri~\cite{sadoc:1999} as well as by Nelson and coworkers~\cite{nelson:2002} is often considered by many to suit this purpose. In addition to encouraging the enumeration of preferred local structures, e.g.,~\cite{miracle:2004}, it has indeed led to the development of a theoretical apparatus for the glass transition based on geometrical frustration~\cite{nelson:2002,tarjus:2005}. Yet, in spite of its marked impact on the structural analysis of dense fluids, this proposal remains largely untested in three-dimensional (3D) systems.

Geometrical frustration is canonically illustrated by considering the behavior of spherical particles of diameter $\sigma$. Because regular simplices (triangles in 2D, tetrahedra in 3D, etc.) are the densest possible local packings of spheres, they are expected to play a central role in liquid organization, e.g.,~\cite{anikeenko:2007,vanmeel:2009}. In 2D Euclidean space, interesting physics results from the fact that simplices can assemble into the triangular lattice~\cite{nelson:2002}, and spatial curvature frustrates the regular assembly of disks~\cite{sausset:2008d}. For Euclidean space in dimension $d\geq 3$ simplices cannot tile space without defects, but in $d=3$ they can form perfect icosahedra on a relatively gently curved space~\cite{foot:1}. The defects that result from uncurving this singular space back to the Euclidean variety can be understood by dimensional analogy. Each particle in a perfect 2D triangular tiling of disks is part of six triangles. Curvature results in irreducible disclinations that sit on disk centers and for which the coordination obtained by a Delaunay decomposition differs from six. Similarly, in 3D each edge between nearest-neighbor pairs is shared by five other tetrahedra; flattening space generates disclination lines of ``bond spindles'' that are shared by $q\neq 5$ tetrahedra. Periodic arrangements of these disclinations form the complex crystal structures known as Frank-Kasper phases~\cite{frank:1959}. Yet even in amorphous configurations \emph{at small frustration}, the simple Voronoi polyhedra that accommodate the presence of spindles with $q\neq5$ provide topological constraints for the propagation of defects from one particle to the next, which results in disclination lines~\cite{nelson:2002}. 
A denser fluid, in which the proportion of $q=5$ spindles grows and conversely that of $q\neq 5$ spindles shrinks, should thus see disclination defects play a more important role. Disclination lines passing one another correspond to activated events, possibly affected by topological constraints~\cite{nelson:2002}; 
the theoretical framework suggests a causality between the dynamical slowdown and a growing static, structural correlation length underlying the fragility of the glass-forming fluid. In this letter, we critically examine this proposal and find that, in spite of its elegance, it does not hold for the system to which it is more directly related, i.e., simple 3D hard spheres. Through a variety of measures of static order, we also consider alternate definitions of correlation lengths and explore in what regime a growing static order could reasonably be associated with a dynamical slowdown in the regime accessible to computer simulations.

\begin{figure}
\includegraphics[width=0.9\columnwidth]{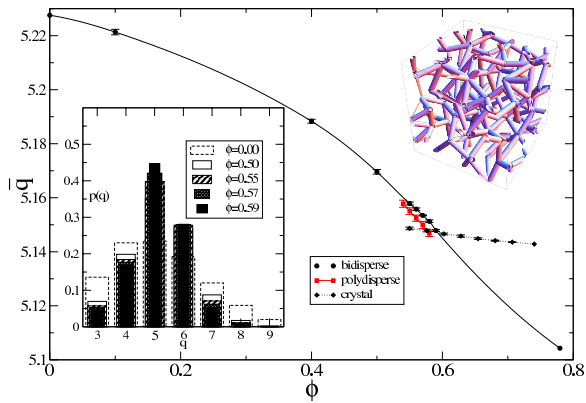}
\caption{Evolution of $\bar{q}$ and (left inset) of the probability distribution $p(q)$ with density. The statistical honeycomb 
limit is given for reference. (right inset) Network of $q=6$ spindles (rods) at $\phi=0.58$.}
\label{fig:spindles}
\end{figure}

Testing this intrinsically geometrical theory on identical 3D hard spheres is problematic because nucleation interferes with the slowdown, so the crystallization drive must be reduced. In case the fluid structure were to sensitively depend on the nature of these perturbations, we do so in two different ways: (i) a 50:50 hard sphere binary mixture with a 1.4:1 diameter ratio whose glass-forming properties have been extensively characterized~\cite{brambilla:2009,flenner:2011,foot:2}, and (ii) a mixture of hard spheres with the smallest non-crystallizing diameter polydispersity, $8.5\%$~\cite{zaccarelli:2009}. The average number of tetrahedra wrapped around a bond, $\bar{q}$, lies within two simple limits. First, all finite-density configurations should have fewer tetrahedra per spindle than a Poisson process (an ideal gas), where $\bar{q}=144\pi^2/(24\pi^2+35)\approx 5.228$~\cite{okabe:2000}. Second, although in curved space the optimal number of simplices per bond can be as low as $q=5$, in Euclidean space a more stringent limit $\bar{q}=2\pi/\arccos{(1/3)}\approx5.104$ is obtained from the fictitious ``statistical honeycomb'' construction
~\cite{coxeter:1961,foot:1}. Figure~\ref{fig:spindles} shows that both densifying fluids present a growing polytetrahedral character. The average spindle coordination decreases, seemingly toward its optimal value, with increasing packing fraction $\phi$ for both models, as does its distribution $p(q)$ (Fig.~\ref{fig:spindles}). The growing simplex order is also quite different from that observed in the face-centered cubic crystal phase. Because the structural properties of the two models are robustly similar, we only consider (i) for the rest of the analysis.

Surprisingly, even for the densest systems equilibrated the disclination network remains highly branched, with multiple defect lines stemming from each vertex. The inset of Fig.~\ref{fig:spindles} illustrates this situation for the $q=6$ spindle network. The typical spacing between defect spindles $\xi_{\mathrm{defect}}\equiv c_{\mathrm{defect}}^{-1/3}$, using a defect concentration
$c_{\mathrm{defect}}\equiv\sum_{q} c_q (q-5)^2$
that puts more weight on higher-order defects, indeed grows by no more than $1-2\%$ over a density range over which the relaxation time goes up by several decades. Extrapolating the results to higher densities using the statistical honeycomb limit further indicates that the growth of $\xi_{\mathrm{defect}}$ remains small over the entire accessible amorphous regime $\phi\lesssim 0.65$. Actually, polytetrahedral order is bound to saturate as a result of the intrinsic frustration of Euclidean space. The saturation length corresponding to the maximal spatial extension of simplex order estimated from the radius of sphere inscribing the gently curved space from which this argument derives 
is $\simeq 1.59\sigma$~\cite{foot:1}. Alternatively, the typical  distance between defects in an ideal tetrahedral structure threaded only by $q=6$ spindles is only $\xi_\mathrm{defect}\simeq 0.99\sigma$, although it is worth noting that the average distance between spindles itself is but $\approx0.3 \sigma$. 
A similar result is obtained for the spatial correlations associated with frustrated local order through an analysis of the spatial decay of the bond-orientational order correlation function $G_6(r)$~\cite{steinhardt:1983,ernst:1991}. No matter how it is precisely defined, the associated correlation length $\xi_6$ does not increase by more than a few percents. Even in 3D, hard spheres are therefore sufficiently frustrated to make the dual picture of amorphous particle packings in terms of spindle defects rather uneconomical at all densities.

\begin{figure}
\includegraphics[width=0.9\columnwidth]{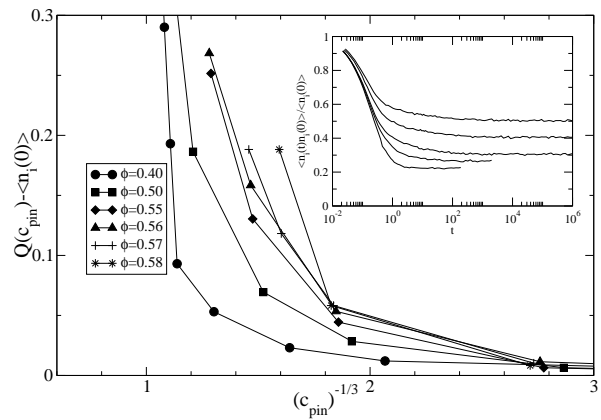}
\caption{Long-time limit of the overlap. (inset) Time $t$ evolution of the overlap for $c_{\mathrm{pin}}=0.11$, 0.45, 0.68, 0.74, and 0.79, from left to right, in $(\beta m\sigma^2)^{-1/2}$ units with mass $m$, the larger particle $\sigma$ and inverse temperature $\beta$ set to unity.}
\label{fig:penetration}
\end{figure}

In order to remove any possible doubt as to whether alternative static lengths due to tetrahedral or other order types are present or not, we turn to the order-agnostic penetration length $\xi_{\mathrm{p}}$~\cite{berthier:2011,cammarota:2011}, which, like the point-to-set length $\xi_{\rm PS}$~\cite{biroli:2008}, characterizes the influence of boundary conditions and is expected to diverge with the relaxation time~\cite{montanari:2006}. It is obtained by pinning a random selection of particles from an equilibrated fluid configuration, and measuring the overlap between the initial and final configurations after a long time $t$ has elapsed,
\begin{equation}
Q(c)=\lim_{t\rightarrow\infty}\frac{\sum_{i}\langle n_i(0)n_i(t)\rangle}{\sum_{i}\langle n_i(0)\rangle},
\end{equation}
where $n_i$ is the occupancy of a spatial cubic cell whose volume is similar to that of the smaller particles in the system in order to prevent multiple occupancy~\cite{berthier:2011}. Subtracting the random overlap contribution $\langle n_i\rangle$ leaves a quantity that grows from low to high as the pinning concentration $c_{\mathrm{pin}}$ increases (Fig.~\ref{fig:penetration});
the crossover is $\xi_\mathrm{p}\sim c_{\mathrm{pin}}^{-1/3}$. Operationally, we define $\xi_\mathrm{p}$ as the value of the average distance for which the overlap falls below $0.18$ (Fig.~\ref{fig:penetration}). The extracted length is not very sensitive to this choice, provided it is intermediate between low and high overlap. We stress that focusing solely on the low-overlap regime provides no information on $\xi_\mathrm{p}$ as it only depends on the standard pair correlation function and therefore on trivial two-point correlation lengths $\xi_2$. This result, which can be checked explicitly by considering the linear response to a vanishingly small $c_{\mathrm{pin}}$, remains true so long as one remains in the low overlap perturbative regime~\cite{charbonneau:2012}. It also casts some doubt on the relevance of a recently proposed scaling~\cite{karmakar:2011}, where the observed linear dependence on concentration suggests instead that only trivial static lengths are probed.

\begin{figure}
\includegraphics[width=0.9\columnwidth]{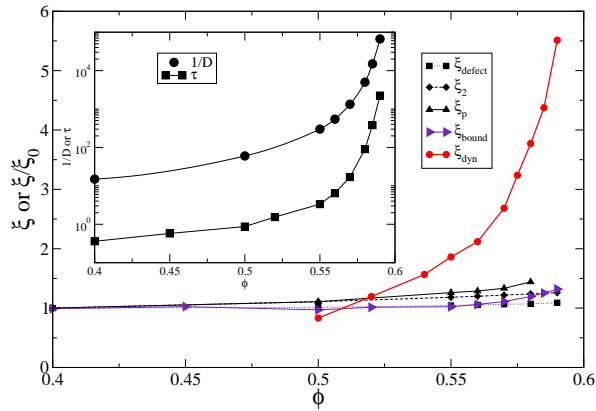}
\caption{Various static lengths rescaled to unity at $\phi=0.40$ ($\xi_0$) together with the lower bound from Eq.~(\ref{eq_bound-length}) $\xi_\mathrm{bound}$ and the dynamic length~\cite{flenner:2011}. (inset) Measures of dynamical slowdown. Lines are guide for the eyes.}
\label{fig:lengths}
\end{figure}

Figure~\ref{fig:lengths} shows that the penetration length $\xi_\mathrm{p}$ increases only very modestly over the dynamically accessible density range. For sake of comparison, we display in Fig.~\ref{fig:lengths} an estimate of $\xi_2$ evaluated from the pair correlation function $g(r)$, which, as is well known, changes only slightly with density. The penetration length increases only about $20\%$ more than $\xi_2$. (Artificially breaking down structural order into radial and orientational contributions suggests that the latter is at most comparable to the former over the density range studied, further supporting the spindle analysis.) Note that in view of the small variation of all the static lengths, $\xi_{\mathrm{defect}}$, $\xi_6$, $\xi_2$ and $\xi_{\mathrm{p}}$, trying to devise a crisper measuring procedure is unnecessary as it will not qualitatively alter the conclusions. Strikingly, the ``dynamic length'' $\xi_\mathrm{dyn}$ characterizing the spatial correlations in the dynamics and associated with dynamical heterogeneities grows markedly over the same density range. Whereas the change in the static length is measured in fractions of their low-density value, $\xi_\mathrm{dyn}$ grows by a factor of almost $7$ when reaching $\phi=0.59$~\cite{flenner:2011}. The diffusivity $D$ and the structural relaxation time $\tau$  meanwhile change by about 4 orders of magnitude (Fig.~\ref{fig:lengths}).

Although the above results may come as no surprise to those who believe the dynamical slowdown to be a purely kinetic phenomenon involving no growing static length scale, it is nonetheless worth checking whether one does not violate the bound between relaxation time and static correlation length put forward by Montanari and Semerjian~\cite{montanari:2006},
$\tau \lesssim \tau_{0}\exp\left (B\, \xi_{\rm PS}^3 \right)$, where $\tau_{0}$ is a constant setting the microscopic time scale. The coefficient $B$ depends on density (or temperature for a glass-forming liquid) and is such that when $\xi_{\rm PS}\sim\sigma$ the right-hand side describes the ``noncooperative dynamics'' of the model~\cite{franz:2011}. Using an Arrhenius-like argument for activation volumes~\cite{berthier:2009}, we note that, all else being equal, higher pressures trivially rescale the free-energy landscape and thereby slow the dynamics. For a hard-sphere fluid, one then expects $B \propto \beta P$ where $P$ is the pressure.  It should be stressed that the upper bound of $\tau$ diverges with the pressure even in the absence of any growing $\xi_{\rm PS}$, as when approaching $T=0$ for an Arrhenius temperature dependence. In the low and moderate density fluid, the relaxation time indeed follows $\tau(\phi)\simeq \tau_{\rm low}(\phi)=\tau_0\exp[K \beta P(\phi)]$ with $K$ a density-independent constant.  One then finds that
\begin{equation}
\label{eq_bound-length}
\left(\frac{\log [\tau(\phi)/\tau_{0}]}{\log[\tau_{\rm low}(\phi)/\tau_0]}\right)^{1/3} \lesssim  \frac{\xi_{\rm PS}(\phi)}{\xi_{\rm PS,0}},
\end{equation}
where $\xi_{\rm PS,0}$ is the low-density limit of $\xi_{\rm PS}$. Equation~(\ref{eq_bound-length}) thus provides a lower bound for the growth of a static length imposed by the dynamical slowdown.

To assess whether the above bound is satisfied or not, one needs an estimate of $\xi_{\rm PS}$. The direct approach would be to consider the effect of pinning the boundary of a spherical cavity on the fluid inside, but one may reasonably expect that the penetration length studied above gives a rough estimate of $\xi_{\rm PS}$. Near a random-first-order transition or near any first-order transition $\xi_{\rm PS}\sim\xi_{\rm p}^3$~\cite{cammarota:2011a}, but far from such transitions, which is the case studied here, one expects $\xi_{\rm PS}\sim \xi_{\rm p}$. In any case, as seen in Fig.~\ref{fig:lengths}, the bound given by Eq.~(\ref{eq_bound-length}) increases only slowly in the dynamically accessible domain and is already satisfied by $\xi_{\rm p}$. Note that this moderate growth of the bound further illustrates that hard spheres are not in fact very ``fragile'' in the regime up to $\phi=0.59$, showing only a limited deviation with respect to the low-density behavior, which is in line with what is found for other simple fluids, such as the Lennard-Jones glass-forming liquids~\cite{grousson:2002}.
These observations may well correlate with the fact that most 3D fluids of spherical particles are strongly frustrated in the sense discussed above.

These results indicate that the growth of a static length is not the controlling factor behind the relaxation slowdown in the range of density considered. This finding points to a mechanism for the slowdown that is either essentially ``noncooperative'', or akin to that predicted by the mode-coupling theory (MCT)~\cite{gotze:2009,biroli:2006}; in both cases, the growth of a dynamic length is not accompanied by that of a static length. We cannot, however, draw any general conclusion on this question beyond this regime. In thermodynamic-based theories~\cite{lubchenko:2007,tarjus:2005}, it is at these higher densities (or lower temperatures for a liquid) where cooperative behavior becomes dominant, and the dynamical slowdown is predominantly controlled by the growth of a static length. This regime is unfortunately mostly beyond present-day computer resources~\cite{foot:3}. A modest indication that a crossover takes place may, however, be given by the data for $\xi_{\rm p}$ and the bound, which both appear to display a steeper increase near $\phi=0.59$.

We have first shown that 3D hard spheres, like $d>3$ hard spheres and many 3D simple glass formers, are too strongly frustrated for the dual picture in terms of defects~\cite{nelson:2002} to bring any useful simplification. One may wonder if there exist other liquids with a different type of locally preferred order for which frustration is weaker and the picture can be put to work. Whereas this frustration regime can be achieved in 2D by curving space~\cite{sausset:2008d}, no such clear-cut example of simulation-accessible glass formers in 3D Euclidean space has yet been devised. Second, we have shown that within the regime accessed here  static lengths grow very slowly, yet in a way that is compatible with the bound recently put forward between relaxation time and static length. For systems similar to hard spheres, these results severely constrain the type of ordering that can develop and place serious doubts on the pertinence of local-order analysis in the moderately viscous dynamical regime. The dynamic length's significant increase points instead to a decoupling between the increasingly heterogeneous character of the dynamics and its cooperative origin in terms of structural or thermodynamic quantities. A challenge would be to search for a possible crossover at still higher densities.

\begin{acknowledgments}
We acknowledge stimulating discussions with G. Biroli, C. Cammarota, D. R. Nelson, R. Mosseri and Z. Nussinov, and thank E. Zaccarelli for sharing configurations. This research was supported in part by the National Science Foundation Grant No.~NSF PHY05-51164. PC acknowledges NSF support No.~NSF DMR-1055586. BC received NSERC funding.
\end{acknowledgments}

\section{Appendix: Geometrical Tilings of simplices}
\newcommand{\R}{\mathbb{R}}
\newcommand{\arccosh}{\mathrm{arccosh}}
We review the regular polytopes and hyperbolic coverings that correspond to defectless tilings of simplices on spaces of constant curvature.  In particular, we aim to show the following.
\begin{enumerate}
\item A regular tiling of tetrahedra is found on the sphere that inscribes the remarkably large 4D platonic polytope $\{3,3,5\}$.  The inscribing sphere's curvature, $\pi^2/25$, is notably much smaller than that of the generalized octahedron $\{3,3,4\}$, $\pi^2/4$.
\item By contrast, on the $d$-dimensional sphere with $d>3$,  $\{3^{d-1},4\}$ is the only way to obtain a regular tiling of simplices resulting in large geometrical frustration of the corresponding Euclidean space, which hints at a possible limit role of simplex ordering in 3D.
\item The regular simplicial tiling $\{3,3,3,6\}$ is found on a hyperbolic space of curvature larger (in absolute value) than the curvature of the sphere that inscribes the tiling $\{3,3,3,4\}$, while the hyperbolic tiling $\{3,3,6\}$'s infinite edge length disqualifies it as a possible reference. 
\item The homogeneous statistical honeycomb $\{3,3,\bar{q}\}$ gives is an upper limit for the density of simplex tiling in $d=3$, and a lower limit to $\bar{q}$.
\end{enumerate}

\subsection{Notations}
We first establish some notation.  The Euclidean space of dimension $d$ is denoted $\R^d$ and the hyperbolic space of dimension $d$ is denoted $H^d$.  The sphere (or hypersphere) of (surface) dimension $d$ is the set of all vectors of length $1$ in $\R^{d+1}$ and is denoted $S^d$.   Hence $S^2$ is the standard radius one sphere in $\R^3$ we all know and love.

Following Schl\"afli's notation (see \cite[Chapter 7]{Coxeter-polytopes}), the polytope $\{a_1,\ldots,a_d\}$ is composed of polytopes $\{a_1,\ldots,a_{d-1}\}$ with vertex figure $\{a_2,\ldots,a_{d}\}$. The \emph{vertex figure} of a $d$-dimensional polytope is a $(d-1)$-dimensional polytope whose vertices are obtained by taking the middle of each edge emanating from a vertex.  This inductive definition deserves an illustration.
The basic example is $\{p\}$, the regular polygon of $p$ sides in $\R^2$.    Next we have the regular polytopes in $\R^3$.  Those are represented by the symbols $\{p,q\}$ for a polytope having faces made of $\{p\}$s and vertex figures made of $\{q\}$s.  Note that because the polytope is regular, one can inscribe all the vertices on a sphere and thus $\{p,q\}$ gives a tiling of the sphere $S^2$. For instance, the cube $\{4,3\}$ is composed of squares (the regular polygon $\{4\}$) and the vertex figure is $\{3\}$, a triangle.   The cubic tiling $\{4,4\}$ in $\R^2$ is composed of 4 squares surrounding each vertex.  

Throughout the text, we use a measure $q$ of the number of tetrahedra sharing a particular edge between nearest-neighbor pairs and $\bar q$ for the average amongst all pairs of nearest-neighbor.  This quantity generalizes to higher-dimensional setting in the following manner: in a $d$-dimensional space (be it $\R^d,S^d$ or $H^d$), a set of  $(i+1)$ points all nearest-neighbors to each other form a $i$-simplex (generalized tetrahedron).  A $1$-simplex is a line segment, a $2$-simplex is a tetrahedron, etc. As in three-dimensional system, the dimension $(d-2)$-simplices play an important role.  We call them \emph{spindles}.  Given a particular $(d-2)$-simplex, we let $q$ measure the number of $d$-simplexes that contain it.   Obviously, $\bar q$ is the average of $q$ over all spindles.

\subsection{Schl\"afli's notation and spindles}
Of primary importance for this paper is the relation between Schl\"afli's notation and the number $q$.  Let $\{a_1,\ldots,a_d\}$ be a regular tiling in $\R^d$ or  $S^d$.  One can show that $q=a_d$. Instead of exhibiting the proof of this result, we give an example. Consider the simple cubic tiling of $\R^d$, given by $\{4,3^{d-2},4\}$.
Here,  a $(d-2)$-dimensional hypercube is given, for example, by $x_1=x_2=0$ and $0\leq x_i\leq 1$ for $3\leq i\leq d$.  The various $d$-dimensional hypercubes containing this $(d-2)$-dimensional hypercube are given by $\pm x_1,\pm x_2\in [0,1]$, where the signs are chosen independently.  Hence $q=4$.

Note that any polytope $\{a_1,\ldots,a_d\}$ in $\R^{d+1}$ can be seen as inscribed in a sphere $S^d$ whose center is the center of mass of the polytope.  By inflating the polytope to make it round, we see that it produces a regular tiling of $S^d$.  For instance, the hypercube $\{4,3^{d-2}\}$ in $\R^{d}$ is also a tiling of $S^{d-1}$.  By looking at the usual case $\{4,3\}$, one sees that it gives $q=3$ on $S^2$.   The same argument illustrates that for the tiling $\{4,3^{d-1}\}$ of $S^{d}$, one also has $q=3$.

\subsection{Various tilings in various spaces}
Consider the case of the generalized octahedron, $\{3^{d-1},4\}$, in $\R^{d+1}$.  It gives a regular tiling of $S^d$ with $q=4$. This tiling is dual in $\R^{d+1}$ to $\{4,3^{d-1}\}$ which can be thought as the cube $[-1,1]^{d+1}$.  So the vertices of the dual generalized octahedron are $\pm \vec{e}_{i}$ for the standard basis $\{\vec{e}_1,\ldots,\vec{e}_{d+1}\}$ of $\R^{d+1}$.  Obviously, the distance between two vertices is ${\sqrt2}$ via straight line, or $\frac{\pi}2$ in $S^d$.  To obtain a unit distance in $S^d$, one has to rescale the sphere to be of radius $\frac 2\pi$, hence the sphere has a curvature $\frac{\pi^2}{4}\simeq 2.47$.

As pointed by \cite[p.20]{sadoc:1999b}, the 120 vertices of the 600-cell $\{3,3,5\}$ all belong to the hypersphere $S^3$ with radius equal to the golden ratio ($\tau=\frac{1+\sqrt5}2$) if the edges are of unit length.  The alternative expression  $\tau=\frac12\csc(\frac\pi{10})$  is useful for our computation.  Indeed, imagine a isosceles  triangle whose equal sides measure $\tau$ and opposite side measure 1.  The angle opposite to the side of measure 1 is \[\theta=2\arcsin\left(\frac 1{2\tau}\right)=2\arcsin\left(\sin(\frac{\pi}{10})\right)=\frac{\pi}{5}.\]
So, if one desires the spherical distance between vertices in the $\{3,3,5\}$ regular tiling of $S^3$ to be 1, the radius must be $\frac{\tau}{{\tau\pi}/{5}}=\frac{5}{\pi}$.  The curvature of this sphere is thus $\frac{\pi^2}{25}\simeq 0.39$.

The tiling $\{3,3,6\}$ in hyperbolic 3-space $H^3$ is somewhat of a degenerate case.  Despite the cells being of finite volume, all of its vertices are at infinity hence the edge length is infinite, which makes the curvature impossible to rescale; see \cite[p.202]{Coxeter-beauty}. In $H^4$ however, the tiling $\{3,3,3,5\}$ is perfectly well-defined, with finite distance between vertices.  According to \cite[p.204 and table p.213 ]{Coxeter-beauty}, the edge-length is $2\phi$ and $\cosh\phi=\tau$, hence the edge-length is $2\arccosh\tau\simeq 2.12$.  The expressions obtained by Coxeter are consistent with the results in standard hyperbolic space of curvature $-1$. Rescaling the space to make the edge-length 1  changes the curvature to $-\bigl(2\arccosh\tau\bigr)^2\simeq -4.505$.

Still following \cite[Chapter 10]{Coxeter-beauty}, one sees that the regular tilings in $H^5$ are not tetrahedral and that there are no regular tilings in hyperbolic space of six or more dimensions. In $H^2$, any $\{3,q\}$ with $q>6$ is possible, with edge-length $2\phi$ where $\cosh\phi=\frac{\cos\frac \pi 3}{\sin \frac \pi q}=\frac{\csc\frac\pi q}{2}$; \cite[p.201]{Coxeter-beauty}. Rescaling to get edge-length $1$, one gets a curvature $-4\arccosh^2(2^{-1}\csc\frac\pi q)$.

The results for $d\geq3$ are summarized in Table~\ref{table:summary}. Note that in the $H^2$ case, the packing of an increasing number of hyperbolic triangles around a vertex, keeping the edge-length 1, makes the curvature go down from approximately $-1.19$ (for $\{3,7\}$), to $-2.34$ (for $\{3,8\}$), to $-3.44$ (for $\{3,9\}$), with limit $-\infty$ as $q\to\infty$.
\begin{table}
\begin{center}
\begin{tabular}{|c|c|c|}
	\hline
	Tetrahedral honeycomb& in & curvature\\
	\hline
	$\{3,6\}$ & $\R^2$&0\\
	$\{3,q\}$ with $q\geq 7$ & $H^2$ & $-4\arccosh^2(2^{-1}\csc\frac\pi q)$\\
	600-cell $\{3,3,5\}$ & $S^3$ & ${\pi^2}/{25}\simeq 0.39$\\
	octahedron $\{3^{n-1},4\}$ & $S^n$ & ${\pi^2}/{4}\simeq 2.47$\\
	$\{3,3,3,6\}$ & $H^4$ & $-4\arccosh^2\tau\simeq -4.505$\\
	\hline
\end{tabular}
\end{center}
\caption{Summary of tiling results with the curvature reported setting $\sigma=1$.}
\label{table:summary}
\end{table}

\subsection{Statistical honeycomb}
Coxeter suggested that nature tries to approximate regular tilings $\{p,3,3\}$ for $p$ between 5 and 6 in soap froth~\cite[Chap. 22]{Coxeter-IntroductionGeometry}.  Obviously, the soap froth itself produces an irregular tiling, but one could look at the average structure, the \emph{statistical honeycomb}.  Taking its dual, as suggested by \cite{nelson:1989b}, we consider the corresponding statistical honeycomb $\{3,3,\bar q\}$ produced by wrapping as many tetrahedra as possible around a given spindle, even if that means wrapping a fraction of a tetrahedron.  Because the angle formed by two adjacent edges in a perfectly regular tetrahedron is $\arccos(1/3)$, the ideal number is thus $\bar q=\frac{2\pi}{\arccos(1/3)}\simeq 5.1043$.  

As proposed in Ref.~\cite{nelson:1989b}, we can consider an ideal configuration where every bond belongs to $\bar q$ tetrahedra.  In the Voronoi decomposition corresponding to this configuration,  the cells have $Z$ identical faces corresponding to the $Z$ nearest neighbors of a particle, each of them being a polygon with $\bar q$ faces.    This Voronoi cell can be seen as living on the surface of a sphere.  If it has $V$ vertices and $E$ edges, we must thus have $V-E+Z=2$, using Euler's relation. Counting the pairs (\emph{edge},\emph{face}), with \emph{edge} adjacent to \emph{face}, in two different ways, we get $2E=Z\bar q$, and counting the pairs (\emph{edge},\emph{vertex}) in two different ways, we get $3V=Z\bar q$.  Overall, we obtain that in this ideal structure, each particle has a number of neighbors
\[Z=\frac{12}{6-\bar q}\simeq 13.4.\]
Note that in this construction, all quantities are also by definition their own average.

Now one can compute the volume occupied by that fictitious Voronoi cell, and obtain, as did Ref.~\cite{nelson:1989b}, a volume of 
\begin{align*}
V_{\text{cell}}&=\frac{Z}{3}\frac{\sigma}{2}\mathrm{area}(\bar q\text{-sided regular polygon})\\
&=\frac{Z\sigma^3\bar q \cot(\frac{\pi}{\bar q})}{144}.
\end{align*}
The packing fraction is then given by 
\[\phi=\frac{V_{\text{sphere}}}{V_{\text{cell}}}=\frac{\frac43\pi(\frac{\sigma}{2})^3}{V_{\text{cell}}}\simeq 0.7796.\]
It should be noted that this result is also Roger's bound on the maximal packing fraction for any sphere packing in 3D, not only simplex tilings.  In all cases, it gives an upper bound on the density of simplex tilings in 3D systems.  


\end{document}